\documentclass[12pt,preprint]{aastex}

\begin{document}


\title{Thirty Meter Telescope Site Testing I: Overview}


\author{M. Sch\"ock,\altaffilmark{1} S. Els,\altaffilmark{2} R. Riddle,\altaffilmark{3} W. Skidmore,\altaffilmark{3} T. Travouillon,\altaffilmark{3} R. Blum,\altaffilmark{4} E. Bustos,\altaffilmark{2} G. Chanan,\altaffilmark{5} S.G. Djorgovski,\altaffilmark{6} P. Gillett,\altaffilmark{3} B. Gregory,\altaffilmark{2} J. Nelson,\altaffilmark{7} A. Ot\'arola,\altaffilmark{3} J. Seguel,\altaffilmark{2}  J. Vasquez,\altaffilmark{2}  A. Walker,\altaffilmark{2} D. Walker\altaffilmark{2} and L. Wang\altaffilmark{3}}

\altaffiltext{1}{TMT Observatory Corporation, NRC Herzberg Institute of Astrophysics, 5071 West Saanich Road
Victoria, BC V9E 2E7, Canada}

\altaffiltext{2}{Cerro Tololo Inter-American Observatory, National Optical Astronomy Observatory, Casilla 603, La Serena, Chile
}

\altaffiltext{3}{TMT Observatory Corporation, 2632 E Washington Blvd, Pasadena, CA 91107, USA}

\altaffiltext{4}{National Optical Astronomy Observatory, 950 N Cherry Ave, Tucson, AZ 85719, USA}

\altaffiltext{5}{Department of Physics and Astronomy, 4129 Frederick Reines Hall, University of California, Irvine, CA 92697-4575, USA}

\altaffiltext{6}{Astronomy Department, California Institute of Technology, MC 105-24,
1200 East California Blvd, Pasadena, CA 91125, USA}

\altaffiltext{7}{Center for Adaptive Optics, University of California, 1156 High St, Santa Cruz, CA 95064, USA}




\begin{abstract}

As part of the conceptual and preliminary design processes of the Thirty Meter Telescope (TMT), the TMT site testing team has spent the last five years measuring the atmospheric properties of five candidate mountains in North and South America with an unprecedented array of instrumentation.  The site testing period was preceded by several years of analyses selecting the five candidates, Cerros Tolar, Armazones and Tolonchar in northern Chile; San Pedro M\'artir in Baja California, Mexico and the 13 North (13N) site on Mauna Kea, Hawaii.  Site testing was concluded by the selection of two remaining sites for further consideration, Armazones and Mauna Kea 13N.  It showed that all five candidates are excellent sites for an extremely large astronomical observatory and that none of the sites stands out as the obvious and only logical choice based on its combined properties.  This is the first article in a series discussing the TMT 
site testing project.

\end{abstract}


\keywords{Astronomical phenomena and seeing, site testing, extremely large telescopes}


\section{TMT site testing and selection basics}

In April 2008, the Thirty Meter Telescope (TMT) Project reduced its short list of candidate sites to two, Cerro Armazones in northern Chile and the ``13 North'' (13N) site on Mauna Kea, Hawaii.  This decision officially ended the TMT site testing work after five years of in-situ measurements, during which at least 2.5 annual cycles of data were acquired on each of the five candidate sites.  The practical work at the sites was preceded by several years of preparatory work, most notably a series of satellite data studies of cloud cover and precipitable water vapor (PWV) of sites in Chile, southwestern North America and Hawaii, on the basis of which the candidate sites were selected.

This paper is the first in a series of twelve articles, hereafter referred to as `TMT Site Testing' 1 to 12 (TST-1 to TST-12), describing the TMT site testing process.  It contains descriptions of the general principles underlying the TMT site testing work and the selection of the candidate sites, as well as summaries of the instrumentation, methodology and the top-level results.  TST-2 \citep{TST-2} provides are detailed account of the process by which the candidate sites were selected.  TST-3 and TST-4 \citep{TST-3, TST-4} are descriptions of the equipment used, the efforts undertaken to ensure data quality and the methods by which the individual pieces were put together to create systems that operate reliably and autonomously at remote sites.  TST-5 to TST-11 \citep{TST-5, TST-6, TST-7, TST-8, TST-9, TST-10, TST-11} contain detailed descriptions of the results obtained with the instrument suite, organized by parameter category: integrated turbulence parameters (TST-5), turbulence profiles (TST-6), turbulence coherence time (TST-7), meteorological parameters (TST-8), cloud cover and light pollution (TST-9), precipitable water vapor (TST-10), and combinations and correlations of parameters (TST-11).  The final paper in the series, TST-12 \citep{TST-12}, describes how this wealth of information was interpreted and used to determine which sites are qualified to host TMT as far as their atmospheric parameters are concerned.

The selection of a site is a critical issue for TMT on many levels.   Obviously, the TMT site needs to be suited for producing astronomical data of superb quality and for maximizing the scientific productivity of the observatory over its lifetime.  In addition, the site has tangible consequences beyond its direct impact on science.  It strongly affects the cost and ease of observatory construction and operation.  It affects the activities of management, technical support, and personnel recruiting.  On a more subtle level, a detailed characterization of the site can affect the telescope and dome design (for example, due to the wind speed distribution, mechanical properties of the soil, seismicity, etc.), and the adaptive optics (AO) design (through the various atmospheric turbulence properties).
Site selection and testing were therefore given high priority from the very beginning of TMT and its precursor projects, the Giant Segmented Mirror Telescope (GSMT), the California Extremely Large Telescope (CELT) and the Very Large Optical Telescope (VLOT).  [For simplicity, we refer to all these efforts as `TMT site testing', even when they happened before the existence of the actual TMT Project.]

Another early decision was not to create requirements for the TMT site in form of limits for certain parameters, as there are generally no hard cut-offs beyond which a site becomes unsuitable. Instead, TMT opted to measure and
predict both the technical and programmatic properties of the sites with the highest accuracy and longest temporal baseline possible within the framework of the program.  The methodology by which these parameters are balanced against each other was developed during the course of the site testing process.  This included input from the different telescope design teams by means of quarterly internal reviews, as well as approximately annual external reviews which were accompanied by the issuance of comprehensive reports of the site testing process and results.  The balancing of the atmospheric parameters is the subject of the last paper of this series, TST-12.  Other technical considerations such as geological and geotechnical conditions of the site (seismic activity, mechanical integrity of the soil, vibration transmission properties, etc.) and programmatic considerations such as construction and operating cost and methods, cultural, environmental and land use considerations, labor force availability,  proximity to astronomers and astronomy infrastructure, the economic impact of siting TMT, permitting, land ownership, availability of infrastructure and transportation, and customs and immigration issues are not discussed in this series as they were not assessed as part of the site testing work.

\section{Selection of TMT candidate sites}

TMT started its site testing and site selection efforts by considering all potentially interesting sites on Earth as potential candidates.  This work started in the late 1990's with a series of meetings and exchanges of ideas and was originally led by the Cerro Tololo Inter-American Observatory (CTIO).  This phase is described in a dedicated paper in this series, TST-2.  In summary, existing information from previous site testing campaigns and from existing observatories was combined with general knowledge of atmospheric behavior.  This process produced the list of `usual suspects' of regions of interest.  For the most part, these can be divided into three types of superb sites in terms of atmospheric properties: 

\begin{enumerate}

\item Coastal mountain ranges next to a cold ocean current with stable subtropical anticyclone conditions. The proximity to the coast allows for unperturbed laminar air flow.  The cold ocean, whose influence may extend to some distance inland, keeps the inversion layer low. These conditions exist on the coasts of California, Baja California, northern Chile and Namibia.

\item Isolated high mountains on islands in temperate oceans, where the weather is good and a 
laminar air flow and large thermal inertia of surrounding ocean keep the inversion layer low.  There are two such locations known on this planet:  the Hawaiian and the Canary Islands, although other locations might exist, such as R\'eunion off the eastern coast of Africa.

\item A number of high points of the Antarctic plateau where katabatic winds and the absence of the jet stream cause low turbulence (above a thin ground layer) and the low temperatures and high altitude create low thermal background and water vapor content.
\end{enumerate}

The first two types of sites have been know for a long time (see, for example, \cite{walker71} and references therein).  The third type has only recently been tested and analyzed quantitatively  (e.g., \cite{lawrence04,swain06}).
While it is generally believed that most inland sites would be inferior to sites in these three categories, mainly due to 
turbulence-generating topography, the situation is not always clear.  An example demonstrating that superb sites (at least in terms of seeing) do not all fit this pattern is Maidanak in Uzbekistan \citep{ehgamberdiev00}.  It is possible that some excellent and so far undocumented sites exist in, for example, northern Mexico, or the high mountains of northern Africa. Site testing in the high Arctic is also beginning to investigate whether sites with comparable conditions can be found there \citep{steinbring08}.

A detailed study of all regions fitting the descriptions above is, of course, impractical even for a project of the magnitude of an extremely large telescope (ELT).  Thus, a first cut, based on both the expected atmospheric properties of the sites and practical concerns, reduced the regions of interest for TMT to northern Chile, the southwestern continental United States, northern Mexico and the Hawaiian Islands.  The TMT candidate site selection process then continued with satellite remote sensing studies of cloud cover and PWV of these regions.  After an initial Chilean satellite survey was completed \citep{erasmus01a}, a second study was undertaken which included the southwest United States and Mexico \citep{erasmus02a}. Finally, a comparison was made between the best sites in Chile and those in the southwest United States plus northern Mexico and Mauna Kea \citep{erasmus03}.  The entire candidate site selection process is described in TST-2.

Using the Erasmus studies as a guide and combining them once more with practical and logistical concerns, three sites in northern Chile, Cerros Tolar, Armazones and Tolonchar; San Pedro M\'artir in Baja California, Mexico and the 13N site on Mauna Kea were identified for further study using in situ measurements.   The elevations and coordinates of the five TMT candidate
sites are listed in Table~\ref{t:candidate_sites}.  
The following section provides general descriptions of the sites and their
locations.

\section{Properties of the TMT candidate sites}
\label{s:candidate-sites}

\begin{table}[!t]
\begin{center}

\begin{tabular}{| l | c | c | c | c | c | ccc |}
\hline
Site Name & Elevation & Latitude & Longitude \\
 & & [deg N] & [deg W] \\
\hline
\hline
Cerro Tolar & 2290~m & -21.9639 & 70.0997 \\
Cerro Armazones & 3064~m & -24.5800 & 70.1833 \\
Cerro Tolonchar & 4480~m & -23.9333 & 67.9750 \\
San Pedro M\'artir & 2830~m & 31.0456 & 115.4691 \\
Mauna Kea 13N & 4050~m & 19.8330 & 155.4810 \\
\hline
\end{tabular}

\caption{\label{t:candidate_sites}List of TMT candidate sites selected for
in-situ testing}

\end{center}
\end{table}

\subsection{Cerro Tolar}

A low elevation site (2290~m) in northern Chile, Cerro Tolar is in the Atacama
desert and has an extremely arid climate.  Tolar is located at a distance of
only 8~km from the coast, at 16~km from the closest paved road and 18~km
north-north-east of Tocopilla, a town of 25,000 inhabitants.  [Note: All
distances given in this section are straight line distances.  Driving distances
are usually 50--100\% longer.]  The closest commercial port, airport
and major population center is Antofagasta (population 225,000), 190~km south
of Tolar.  There is a primitive four-wheel drive (4WD) road to the summit, where some
radio equipment is installed.  

The summit area is small and a significant amount of
earth would have to be moved to accommodate TMT.  We are not aware of Tolar
having any particular cultural or archaeological significance to the local people and communities.

In spite of its closeness to Tocopilla, light pollution is not an issue,
as the bluffs above Tocopilla are $\sim$1000~m high and block most of the light
produced in town, reducing light pollution to a faint glow close to the horizon.  Some small light sources from mines and a train station
are visible in the south and south-east.

\subsection{Cerro Armazones}  

Cerro Armazones, a medium elevation site (3064~m) in northern Chile,  is also
located in the Atacama desert and close to the coast (36~km), with a climate
very similar to that of Tolar.  It is 22~km from ESO's Very Large Telescope
(VLT) on Cerro Paranal (2635~m) and 110~km south of Antofagasta, the closest city.   A
good, but steep and narrow switch-back road to the summit exists.  The closest
paved road is $\sim$18~km from Armazones, connected by a rough dirt road.

Armazones is the site of a small observatory operated by the Universidad
Cat\'olica del Norte in Antofagasta.  This observatory is not located on the
summit, but on a saddle $\sim$350~m below the summit.  A new observatory
utilizing a hexapod mounted telescope \citep{chini00} is being commissioned by the University Bochum
on Cerro Murphy, a small peak 1.5~km south-west of Armazones and $\sim$225~m
lower.  

The summit area is small, albeit somewhat larger than that of Tolar, requiring
leveling down to $\sim$12~m below the current high point for TMT to be built on
Armazones.  No particular cultural or archaeological significance of Armazones is known. 
An archaeological study of the mountain found no artifacts \citep{armazones_archaeology_study}.

The only lights visible from Armazones are glows close to the horizon from Antofagasta in the north-north-west and from large, albeit distant mines, in particular to the east-north-east.  Some prospecting is going on
in the area and needs to be monitored.

\subsection{Cerro Tolonchar}

Cerro Tolonchar is the eastern-most of the Chilean sites, south of the Salar de
Atacama, and only 25--80~km from several 5000--6000~m peaks of the Andes. 
Because of its eastern location and higher altitude, it experiences more
precipitation and clouds than Tolar and Armazones, especially during the South American summer monsoon \citep{zhou98}, also known as the
``Invierno Altiplanico (Altiplano Winter)'', from approximately mid December to mid February.  Tolonchar
is also the highest (4480~m) and most remote of all TMT candidate sites.  The
closest villages are Peine and Socaire (both approximately 300 inhabitants) 30--40~km to the north, with
Toconao (550 inhabitants) at 80~km, San Pedro de Atacama
(an eco-tourism town of 1,500 people) at 115~km and Calama, the next large city
with a commercial airport (120,000 inhabitants), at 190~km.  The driving time
is currently 2~h from Socaire, 3~h from San Pedro de Atacama and 4.5~h from
Calama.  Antofagasta, 250~km distant, can be reached via a different route in
$\sim$5.5~h.  These times can be reduced by 30--60~min through the construction
of a good road to Tolonchar.  Currently, only a rough 4WD road exists
from the Paso Sico road to the base of Tolonchar (closest distance $\sim$17~km).  TMT has constructed a road
from the base to the summit which is designed to be usable, with some
improvements, for the observatory.

The summit area is large and flat and would require little work to accommodate
TMT.  There is a stone structure of cultural significance on the summit
and some artifacts were found, and carefully avoided, during the road
construction.  Given the size of the summit area, it should be possible to
avoid any such structure even for a building the size of TMT.  Tolonchar has
significance to the local people and communities.  This is under investigation.

Some lights of mines in and around the Salar de Atacama, approximately 50~km distant but
with a direct line of sight, from the large Mina Escondida in the south-west and from the towns described above are visible close to the horizon,
but Tolonchar remains a very dark site.

\subsection{San Pedro M\'artir}

San Pedro M\'artir (SPM) is located in northern Baja California, Mexico, inside
a national park and is the site of the Observatorio Astron\'omico Nacional de
San Pedro M\'artir \citep{lopez03}.  It is a medium-elevation site (2830~m), $\sim$65~km from
the Pacific coast in the west and 55~km from the Sea of Cortez (Gulf of
California) to the east.  The terrain is gently rising from the north, west and
south, with a steep cliff dropping more than 2000~m to the desert in the east. 
The highest point of the area and, in fact, of Baja California, Picacho del
Diablo (3095~m), is approximately 6~km to the south-east of the observatory. 
The area is inside a pine forest and receives more precipitation than the Chilean
TMT candidate sites, although most of that comes down in a number of
strong events with mostly clear time in between.  The closest town is
Ensenada (300,000 inhabitants) at 4~h driving time and 140~km line-of-sight
distance.  The closest commercial airports are in Tijuana (at 220~km) and San
Diego (250~km).

There is an existing road all the way to the observatory.  It is paved to the
national park ranger station, $\sim$8~km from the observatory, after which it would
have to be improved for an operation of the magnitude of TMT.  Some work in the summit area,
potentially involving moving one of the existing telescopes, would be required
to accommodate a building of the size of the TMT observatory.

The surrounding area is very dark from the north-east through the south to the
west.  In northern directions, the San Diego/Tijuana and Mexicali/Yuma
(180/200~km) areas and Ensenada in the north-west produce visible glows, but
due to their distance, San Pedro M\'artir remains a very dark site.

\subsection{Mauna Kea 13N}

The TMT candidate site on Mauna Kea on the Big Island of Hawaii is a location
referred to as ``13 North'' (13N) on the northern shield, approximately 150~m
below the summit.  It is adjacent to the Submillimeter Array (SMA) extension
area.  With $\sim$4050~m elevation, 13N is the second highest of the TMT
candidate sites.  The conditions are usually dominated by a stable
north-easterly flow, but can produce severe weather and precipitation, in
particular in the winter. As a developed site with several observatories, much
of the infrastructure required for TMT exists on Mauna Kea.  Only a short piece
of road would have to be constructed to the 13N site.  The 13N area is
relatively flat, but some earth moving would be required nevertheless due to
its location inside a somewhat sloping lava field.

Mauna Kea is of great cultural and archaeological significance to the local
people.  What effect this has on the potential construction of TMT at 13N is
currently under investigation.

The lights from most of the towns in the north and west of the Big Island are
visible from the 13N location, as well as the glow from Hilo (at 45 km distance; population 40,000).  However, as for all the other sites, the vertical extent of the light pollution remains well below the 65 degree zenith angle observing limit of TMT [see TST-9 for details on the light pollution studies].

\section{The TMT site testing instrument suite and methodology}
\label{s:instruments}

The top level requirement for TMT site testing was to produce data sets that can be compared quantitatively with the highest possible level of confidence and with known uncertainties.  It was decided from the beginning that this can only be achieved by using identical sets of equipment on all sites under operations conditions that are as identical as is feasible.  A large effort was spent on calibration and comparison of instruments, including side-by-side comparisons of identical instruments and of different instruments measuring overlapping parameter spaces, sensitivity analyses of the dependence of the results on input and calibration parameters, as well as on independent verification of all in-house analysis software by at least two people and independent verification of all results and statistics by at least two people (see, for example, \cite{wang07, els08_apopt, travouillon06}).

The entire TMT site selection instrument suite and its calibration and verification is described in TST-3 and TST-4.  Details of importance for the data analyses and interpretations of results from specific instruments as well as about the instruments themselves, along with references, are given in TST-5 -- TST-11.  This section only summarizes the atmospheric parameters of interest for TMT, the instrumentation used at the TMT candidate sites and the approach used in the interpretation of the measurements.

The TMT Science-based Requirements Document (SRD) requires the TMT site to enable maximum use of TMT as a facility planned to
operate in the 0.3 to 30 $\mu$m wavelength range with adaptive optics as an
integral element in achieving the specified performance.  Among the key scientific and technical features listed as desired in the SRD are a high fraction of clear nights, excellent image quality (seeing), large isoplanatic angle, long turbulence coherence time, small outer scale, low precipitable water vapor, low typical temperatures, low wind speed distribution to limit telescope buffeting but sufficiently high wind speed to enable enclosure flushing, minimal change of temperature during the night, minimal seasonal temperature variations and minimal day-night temperature variations.  Also listed as desired are high altitude, which is a factor generally creating favorable conditions for some of the previous characteristics rather than being a requirement by itself, and a site latitude that creates overlap with observatories such as the Atacama Large Millimeter/Submillimeter Array (ALMA), which is considered important to achieve some of the TMT science goals.  

From the science-based requirements as well as other TMT design requirements, the TMT site testing group developed a list of atmospheric parameters which, ideally, should be measured at each site:
\smallskip

\noindent Weather-related characteristics:\\
\hspace*{1cm} Fraction of cloud cover\\
\hspace*{1cm} Fraction of photometric conditions\\
\hspace*{1cm} Low-elevation wind profile (below $\sim$800m)\\
\hspace*{1cm} High-elevation wind profile ($\sim$800m and above)\\
\hspace*{1cm} Air temperature at several elevations above ground and soil temperature\\
\hspace*{1cm} Ground-level humidity\\
\hspace*{1cm} Precipitable water vapor\\
Turbulence-related characteristics:\\
\hspace*{1cm} Overall seeing\\
\hspace*{1cm} Turbulence strength profiles (through the entire atmosphere)\\
\hspace*{1cm} Isoplanatic angle\\
\hspace*{1cm} Turbulence time constant\\
\hspace*{1cm} Outer scale of turbulence\\
Other characteristics:\\
\hspace*{1cm} Low-elevation dust concentration\\
\hspace*{1cm} High-elevation dust\\
\hspace*{1cm} Light pollution\\
\hspace*{1cm} Atmospheric transparency\\
\hspace*{1cm} Sky brightness\\
\hspace*{1cm} Sodium layer properties\\

For practical reasons, not all of these parameters could be measured during the TMT site
selection process as this would have been beyond the means
of the project.  Specifically, these are the outer scale of turbulence, high-elevation dust, sky brightness and sodium layer properties.  The atmospheric transparency was not measured quantitatively.  The high elevation wind profile could also not be measured, but was estimated from radiosonde (balloon) and National Centers for Environmental Prediction (NCEP) reanalysis data.   Nevertheless, the majority of parameters identified above were measured with on-site equipment for periods of time spanning from 2.5 to 5 years at each of the five candidate sites.

The following instruments have been deployed at the candidate sites:
\begin{itemize}

\item {\bf  Differential Image Motion Monitors (DIMM):} The TMT DIMMs are
mounted on small (35~cm) but robust custom-made telescopes installed on 6.5~m
towers.  A DIMM measures the integrated seeing in the air column above the
telescope \citep{sarazin90,wang07}.  It can also be used to obtain estimates of the isoplanatic angle, turbulence coherence time and cloud cover and transparency along the line of sight.  However, as other instruments measure these latter parameters with higher confidence, we only use the DIMM seeing measurements for TMT site testing and selection purposes.

\item {\bf  Multi-Aperture Scintillation Sensors (MASS):} The MASS is a scintillation-based
instrument which measures six-layer turbulence profiles excluding the ground layer, the isoplanatic angle
and the turbulence coherence time \citep{kornilov03,els08_apopt}. We also use the TMT MASS data for atmospheric
transparency estimates along the line of sight.

\item {\bf  Sound Detection and Ranging (SODAR) acoustic sounders:}
The SODARs used for TMT site testing are phased-array acoustic emitter/receiver systems which produce low elevation  turbulence and wind profiles \citep{travouillon06}.  Two models, denoted XFAS (40~--~800~m range, 20~m resolution) and SFAS (10~--~200~m range, 5~m resolution), are used.

\item {\bf  Automatic Weather Stations (AWS):} These are commercial weather stations with
temperature (air and soil), wind speed and direction, humidity, barometric
pressure, precipitation, solar irradiation, ground heat flux and net radiation
sensors. Stand-alone units are mounted $\sim$2~m above the ground. Air temperature
sensors are also installed on 30~m towers on Armazones and Tolonchar.

\item {\bf  Sonic Anemometers:} Mounted at the MASS/DIMM telescope level and/or
at several elevations on the 30~m towers, sonic anemometers measure wind speed and direction and an approximate temperature value, and can be used to estimate the in-situ
turbulence strength.  [During their site survey, the Large Synoptic Survey
Telescope, LSST, \citep{ivezic08} project also had a 30~m tower with sonic anemometers installed
at San Pedro M\'artir.  The data from this are available to us.]

\item {\bf  All-sky cameras (ASCA):} The TMT ASCAs provide images of the entire sky in
several visible and infrared filters \citep{walker06}. They are used for cloud analyses and light pollution studies.  [Note: The ASCA at San Pedro M\'artir was also owned by LSST.]

\item {\bf  Infrared Radiometers for Millimetre Astronomy (IRMA):} An IRMA measures the
flux from the sky at 20~$\mu$m. The precipitable water vapor (PWV) content of
the atmosphere can be calculated from this using a suitable atmospheric model \citep{chapman04}.

\item {\bf  Dust Sensors:} Commercial particle counters are mounted at the
MASS/DIMM telescope level. They measure the particle density in five different
channels for particle sizes of 0.3, 0.5, 1.0, 2.0 and 5.0~$\mu$m.

\end{itemize}

Note that all wavelength-dependent turbulence parameters given in this paper are calculated for a wavelength of 0.5~$\mu$m.  All values except the turbulence coherence time, $\tau_0$, are also corrected for the direction of observation and refer to zero zenith angle.  The correction of $\tau_0$ for zenith angle depends on the high elevation wind direction which is not known.  The consequences for the interpretation of the $\tau_0$  results are explained in TST-7, where we show that the errors introduced by this lack of correction are on the order of the inherent uncertainties of the $\tau_0$ measurements.

One of the key challenges of any site testing campaign is the fact that atmospheric and climate variations exist on all time scales.  Thus, a site that shows excellent conditions during site testing might later experience poorer conditions, or vice versa.  This is unavoidable, but measuring for the longest period possible obviously increases the confidence that conditions during the site testing period are representative of the long term properties of the site. 
The original goal of the TMT site selection campaign was to take on-site
measurements of the most important parameters (e.g.\ weather, seeing) for at least two
annual cycles, and for at least one year for all other parameters. This was achieved or
exceeded, in some cases significantly, for most instruments, but was not possible in all cases for practical
reasons.  Note also that we only have three sets of SODARs and IRMA water vapor radiometers.  These instruments have therefore not been installed continuously at all sites, and have not operated at all at some sites.  Details about the deployment schedules and the amount of data available from each instrument are given in the other TST papers.

Inspection of the multi-year records of atmospheric conditions at the TMT candidate sites shows that most have stable or repeating patterns throughout the years.  However, we also observe months and seasons that show significantly different conditions from one year to the next for certain parameters at some sites.  With generally ``only'' two to four years of data available, it is not always possible to say which of these periods are typical, or more typical, for long-term conditions. This is, however, a very important question for an observatory project with an expected life time of 50 years.

We attempted to address the representativeness of our data by comparing them with other, longer-term data sets.  First, for some parameters, we analyze long-term records available from satellite and climate data.  The satellite data directly produce estimates of cloud cover and precipitable water vapor such as those used in the selection of the candidate sites.  While the climate data do not provide direct measurements of the parameters of interest for the TMT candidate sites, they can be searched for changes in key climate parameters that might affect the conditions at the sites and indicate non-representative periods.  See TST-9 and TST-10 for details.

Second, San Pedro M\'artir and Mauna Kea are sites of existing observatories which,
over the years, have seen many site characterization efforts [for example, \cite{cruzgonzales03,erasmus86,tokovinin05}].   Data on image
quality and environmental conditions from the observatories themselves are also
available.  The closeness of Armazones to Paranal might make some
of the large amount of data from the VLT Astronomical Site Monitor applicable
to Armazones.  It is therefore tempting to use these data to extend the temporal baseline of the measurements available for the TMT candidate sites.

It is, however, important to recognize that a reliable
comparison of data sets is only possible if great care is taken to calibrate all
instruments, stringent setup and operation procedures are adhered to and analysis
criteria are applied in equivalent ways.  One of the main lessons learned during the TMT site testing work is how difficult it is to obtain high-accuracy (or even high-precision) data even with identical equipment, in particular for turbulence measurements.  The comparison
of results from non-identical instruments that have not undergone the same kind
of stringent calibration and data control procedures can easily lead to
misinterpretations of the results. This is the main reason why the TMT site
selection project put such importance on the use of identical equipment at all
candidate sites and on rigorous characterization of our instrument suite.  The bottom line is that a straight comparison of our results with those from other campaigns
is, in most cases, not meaningful and external results are generally not used in the TMT site selection process.  

We nevertheless made every effort possible to verify that our results are consistent with data taken from other sources.  This is, in particular, the case for seeing, wind and PWV measurements.
When overlapping data sets are available, we generally find that our data are consistent with those from other sources within the uncertainties of the comparisons, but that those uncertainties are often too large to be useful for TMT site selection purposes.  Where applicable, these efforts are described in the other papers of the TST series.

\section{TMT site selection top-level results}
\label{s:summary-results}

This section provides a top-level summary of the results from the TMT candidate sites, with many more details given in TST-5 -- TST-11. It is obviously impossible to present all aspects of the huge amount of data collected during the TMT site testing work in one paper, or even in twelve.  There are simply too many different properties of that data that might all be interesting for one or another purpose.  In addition, there are generally also a multitude of different ways of analyzing and interpreting the same subset of data.  One example of this is the seemingly simple compilation of the DIMM seeing probability distributions.  The most straight-forward method to do this is to weight all individual data points equally.  Other methods might involve the calculation of monthly (or seasonal) distributions and averaging those, either with equal weights or weighted by the length of night time and the weather conditions during different parts of the year.  All these methods have their advantages and disadvantages depending, for example, on what the results are to be used for, how evenly the data cover each month or season of the year,  or whether expected up and down times of the observatory are known as a function of the month/season.

A second example of different interpretations of the same data set is the exclusion of invalid data points, such as times when a mechanical anemometer gives zero wind speed readings because it is frozen over after a winter storm.  Obviously, such points should be excluded from the analysis.  However, unless an {\it independent} record exists for when this happened (which usually is not the case for a remote site), it is difficult to identify and exclude these and only these events.  Anemometers also read zero wind speed in normal operation, when the wind speed is below their detection threshold.  This threshold increases in time as dust gets into the gears and lubrication decreases.  Thus, excluding all zero wind speed readings might bias the statistics more than leaving frozen-sensor data in the set.  Which way of data analysis is preferable depends, again, on its goal (for example, whether the high or low wind speed regime is of primary interest) and on which conditions exist more frequently at a given site.

These examples illustrate that there is not one single way of representing the TMT site testing results that is applicable to or useful for all purposes.    
Unless noted otherwise, the probability distributions presented in the papers of the TST series are therefore generally simply the distributions of all data taken during the site testing campaign, excluding only data points that are absolutely certain to be invalid.  This does not mean that, over the last five years, we have not looked at many other aspects of the data sets.  Most of the time, we find that the differences between analysis methods are small enough that they will not make a difference for the purpose of TMT site selection (but they might be important for some other application).  If significant differences are found, we make sure that we understand their causes and consequences and use the most appropriate analysis method.  In some cases, these different approaches are described in TST-5 to TST-11.  However, an all-inclusive description of the TMT candidate sites is beyond the means of any publication.  Instead, TMT will make its site testing data base public after the publication of this paper series to allow individual, specialized analyses by the interested reader.

\begin{figure}[!t]
\resizebox{1.0 \textwidth}{!}{\rotatebox{-90}{\includegraphics{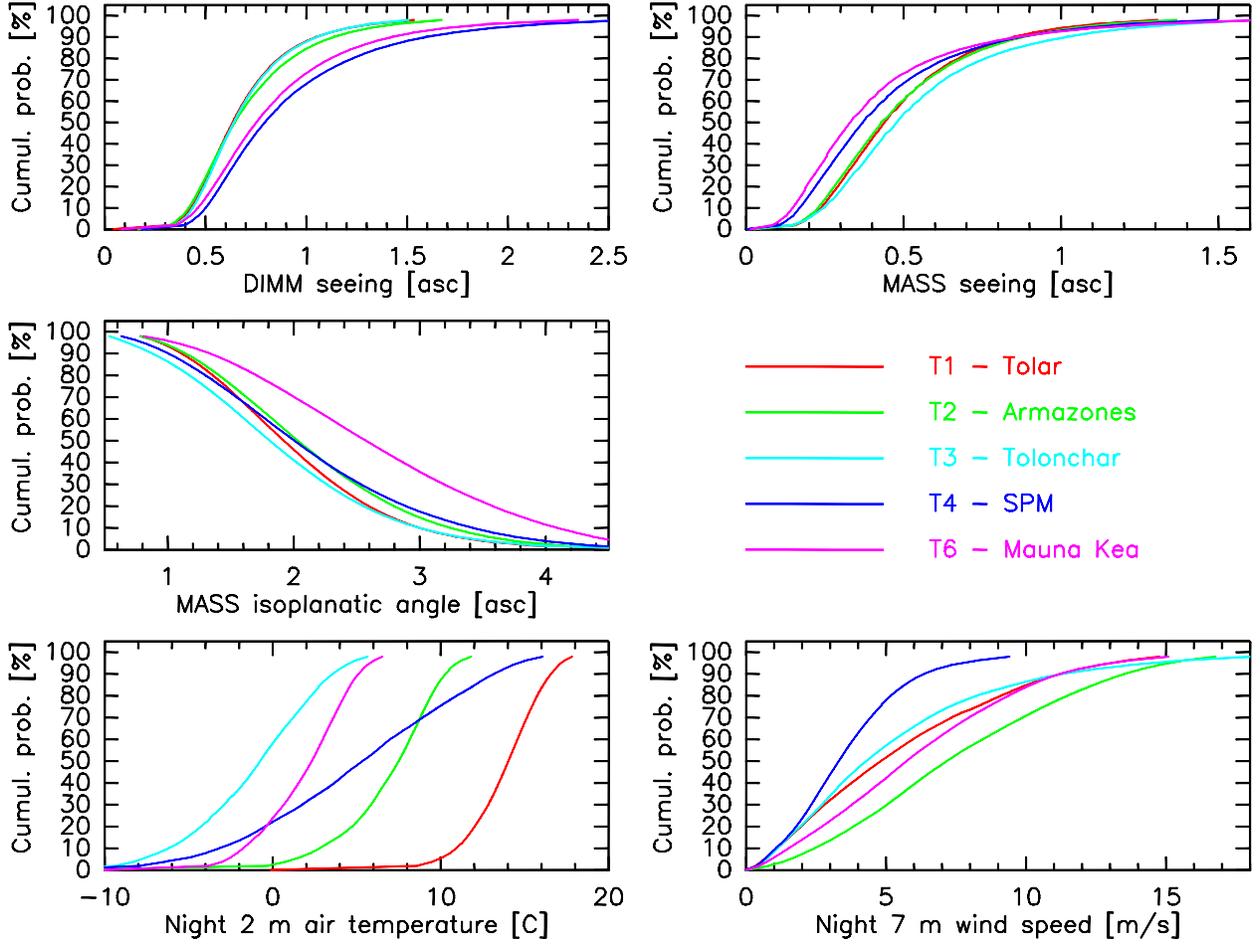}}}

\caption{\label{f:cumulcolor}Cumulative probability distributions for the five
candidate sites.  DIMM and MASS seeing, isoplanatic angle, temperature and
wind speed are shown.}

\end{figure}

\begin{figure}[!t]
\resizebox{1.0 \textwidth}{!}{\rotatebox{-90}{\includegraphics{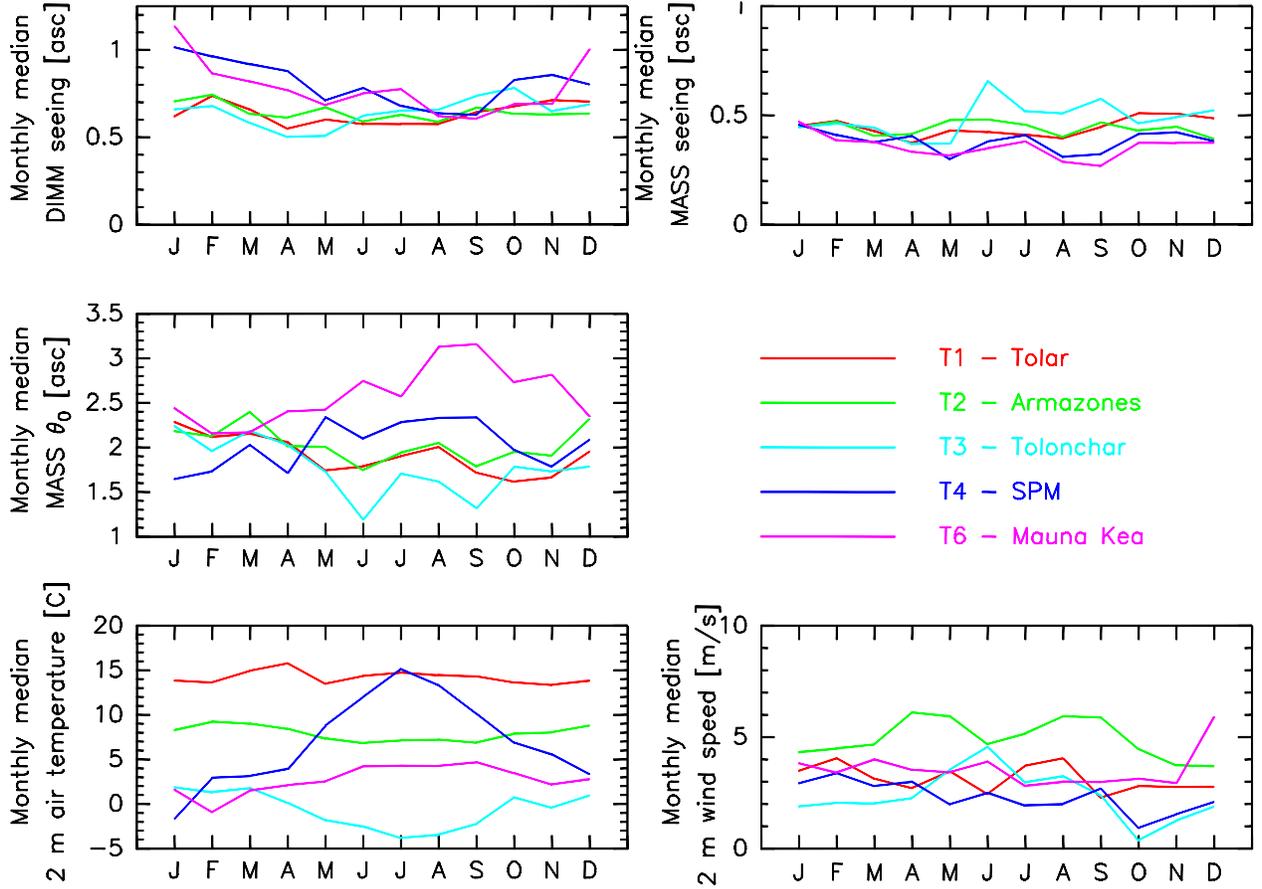}}}

\caption{\label{f:monthlycolor}Monthly median values for the five candidate
sites.  The DIMM and MASS seeing, isoplanatic angle, temperature and wind speed
are shown.  The AWS 2~m
wind speeds are used here instead of the 7~m sonic anemometer measurements
because the AWSs have been operating at the sites for much longer than the
sonics.  Note that not all the
ordinates of the graphs start at zero.}

\end{figure}

\begin{figure}[!t]
\resizebox{0.8 \textwidth}{!}{\rotatebox{0}{\includegraphics{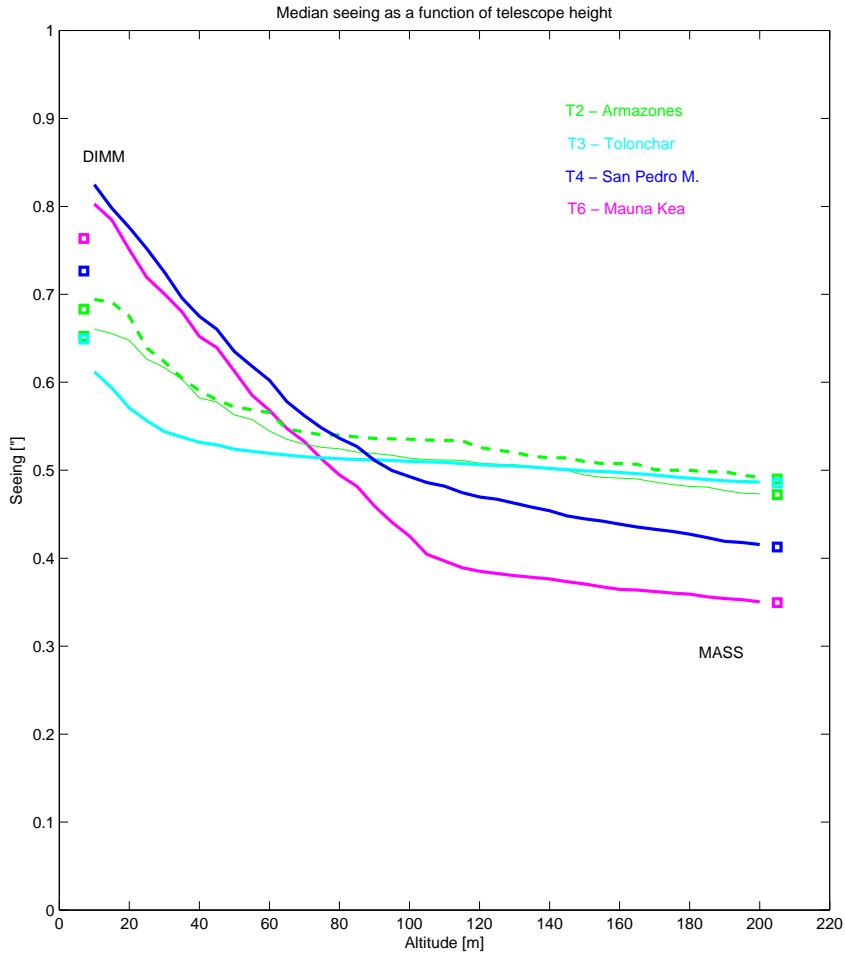}}}

\caption{\label{f:seeing_height}Median seeing an observer would experience at a
given altitude above the ground as calculated from the MASS, SODAR and DIMM
turbulence measurements (simultaneous data only), from 7 to 200~m above the ground.  Note that the simultaneous data cover much shorter periods of time than the overall site testing period, which means that these data should not be used to
compare the absolute magnitudes of the ground layer seeing between the sites. 
They can, however, be used to get a general feeling for the shape of the ground
layer profiles at the sites.  See text for details.}

\end{figure}

\begin{figure}[!t]
\resizebox{0.7 \textwidth}{!}{\rotatebox{0}{\includegraphics{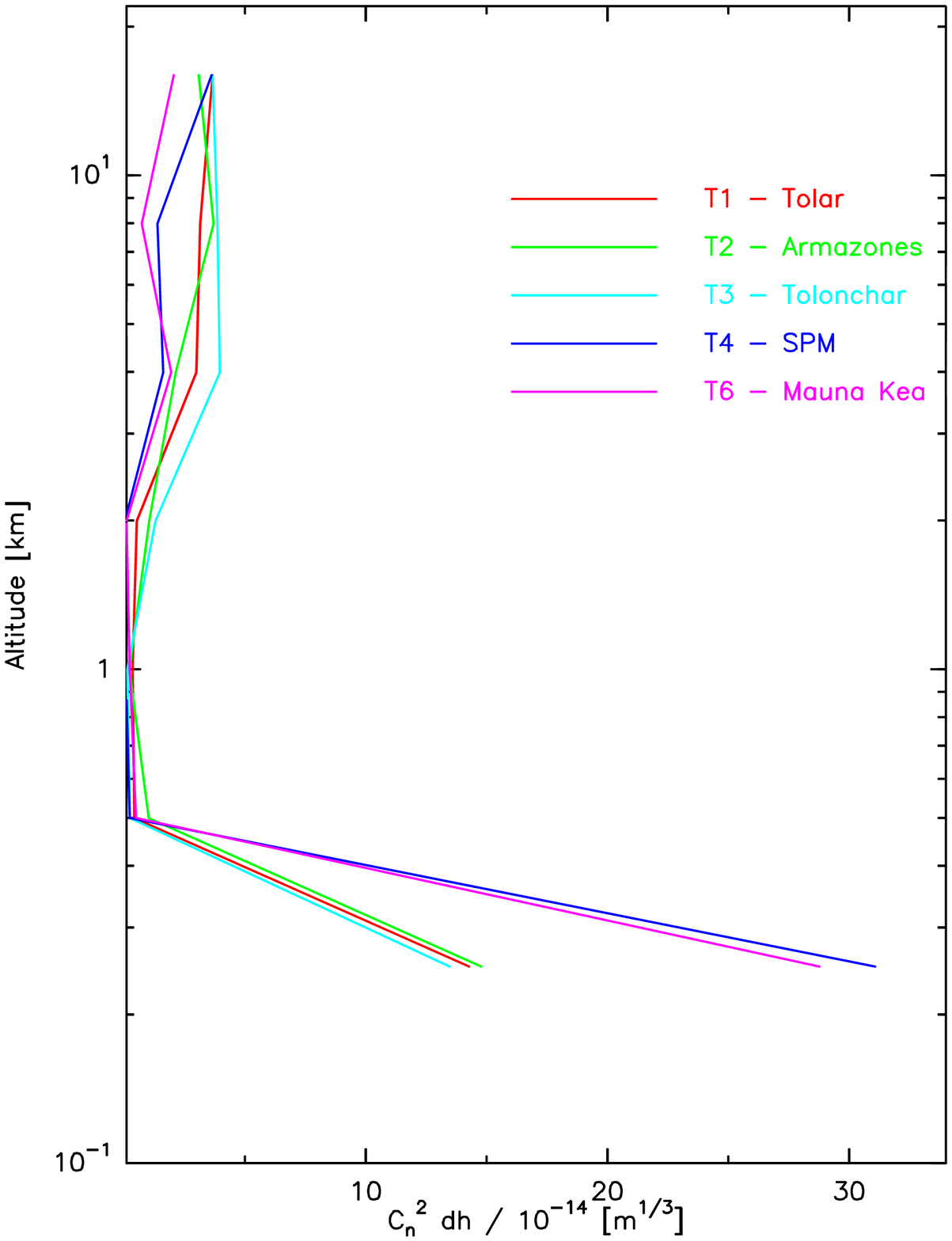}}}

\caption{\label{f:cn2color}Median turbulence strength ($C_n^2 {\rm d}h$)
profiles for the five candidate sites.  The top six levels are the MASS
profiles, while the lowest point (plotted here at 250~m) is the ground layer
strength as calculated from the difference between the DIMM and the MASS
seeing.}

\end{figure}

Plots comparing the main characteristics of the TMT candidate sites are
presented in Figs.\ \ref{f:cumulcolor} -- \ref{f:cn2color}. 
Figure~\ref{f:cumulcolor} shows the cumulative probability distributions for
each site for DIMM and MASS seeing, MASS isoplanatic angle, night-time air
temperature at 2~m above the ground and night-time wind speed at 7~m.  It can
be seen that the total seeing, as measured by the DIMM, is smaller at the
Chilean sites than at the North American sites, with Tolar, Armazones and
Tolonchar being almost identical.  San Pedro M\'artir has a slightly larger
total seeing at 7~m above the ground than Mauna Kea 13N.  

For the MASS seeing (the seeing integrated from $\sim$500~m to the top of the
atmosphere), the situation is reversed, with Mauna Kea 13N showing the weakest
high-altitude turbulence, followed by San Pedro M\'artir, Tolar/Armazones and
Tolonchar.  This shows that the larger DIMM seeing of the North American sites
comes from their having a stronger ground layer than the Chilean sites (see
Fig.~\ref{f:seeing_height} and discussion thereof for more details).  

The isoplanatic angle behavior is similar to that of the MASS seeing, with
the exception that San Pedro M\'artir does not have an isoplanatic angle
significantly better than the Chilean sites.   The high-elevation turbulence
profile at San Pedro M\'artir, while having a lower integrated value than the
Chilean sites, has a highest (16~km) layer which is approximately equally as
strong as those of the Chilean sites (see discussion of Fig.~\ref{f:cn2color}).  As the
isoplanatic angle is dominated by high altitude turbulence, this causes the isoplanatic
angle of San Pedro M\'artir to be comparable to that of the Chilean sites.

The shapes of the temperature distributions in Fig.~\ref{f:cumulcolor} are
similar for Tolar, Armazones, Tolonchar and Mauna Kea 13N, with the slightly
flatter shape of the Tolonchar curve indicating a little more seasonal
variability than the other three sites (see also Fig.~\ref{f:monthlycolor}). 
The median values are a reflection of the site altitudes, Tolar being the
lowest and warmest and Tolonchar being the highest and coldest.  The median
value of the San Pedro M\'artir temperature also follows approximately this
trend in altitude (it is somewhat colder than Armazones, which is of similar
height), while the shape of the curve indicates a larger annual
temperature range.  This stems from the fact that temperatures at San Pedro
M\'artir show larger seasonal variations than at the other sites (see Fig.~\ref{f:monthlycolor}).

Finally, the wind speed distributions (sonic anemometer wind speeds at 7~m)
show Armazones as the windiest site, followed by Mauna Kea 13N and Tolar. 
Tolonchar has surprisingly low wind speeds for a high-elevation site.  Note
that the free-air wind speeds at San Pedro M\'artir are higher than indicated
by our measurements which are affected by the presence of trees at the site. 
This effect is more significant for the 2~m AWS wind speed sensor than for the
7~m sonic anemometer, but it is not negligible for either instrument.  Based on
data from the 30~m tower at San Pedro M\'artir and on existing
data from the observatories, it appears that the nighttime wind speed measured
by the sonic anemometer underestimates the free air flow by a factor 1.5~--~2, which makes the wind speeds comparable to Tolonchar, Tolar and Mauna Kea \citep{michel03}.

Also note that, while the 7~m wind speed at Armazones is higher than at the other sites, its wind speed profile is essentially flat above that.  This means that the wind speed differences between Armazones and the other sites at the level of the TMT enclosure openings are smaller than at the 2~m and 7~m levels.

The monthly median values of the DIMM and MASS seeing, MASS
isoplanatic angle, nighttime air temperature and
nighttime wind speed at 2~m are given in Fig.~\ref{f:monthlycolor}.  The results are represented in a ``standard year'' fashion, meaning that all data taken in a given month are averaged regardless of the year in which they were taken.  Note that
the interpretation of variations in this figure requires the consideration
of the number of data points available for each month and the overall length of time for which data are available.  A month, or even season, might appear different from the rest of the year because the weather was unusual for one season and only two or three years of data are available.  This is one of the inherent problems with limited data sets as described above.  Small variations in the monthly medians should therefore be taken with care.  Nevertheless, some trends (or the absence thereof) can be seen in Fig.~\ref{f:monthlycolor}.   More information is available in TST-5 to TST-11 as well as in the plots for the individual sites in the final TMT Site Testing Report, which will be made public in the future.

The DIMM seeing of Tolar and Armazones displays little
seasonal variability, with some variability at Tolonchar and the North American sites, where it appears to be best in late summer and early fall.  The behavior is similar for the MASS seeing, although the differences in monthly MASS seeing are smaller than for the DIMM seeing for the North American sites, indicating that most of the variations there happen in the ground layer.
There seems to be some variability of the isoplanatic angle at all sites, with the largest (best) isoplanatic angles happening in the summer.  The variations are largest for Mauna Kea, followed by Tolonchar.
There is little seasonal variation of the average
temperature for Tolar and Armazones, with some variation for Mauna Kea 13N and
Tolonchar.  San Pedro M\'artir shows the strong seasonal temperature variability
noted above.  There is some evidence for small seasonal variations of the wind
speeds for most of the sites, in particular for Armazones.  As an aside, we point out that the winter of 2007 was harsher (significantly more windy and somewhat colder) at the Chilean sites than the other winters for which we have data.

The results of the ground layer measurements from the SODARs are shown in
Fig.~\ref{f:seeing_height}.  The quantity plotted here is the integrated seeing
encountered by an observer at a given height above the ground arising from all turbulence components at and above the altitude of the observer.  The square on
the right is the median MASS seeing (the seeing from approximately 500~m
up; plotted here at the 205~m level, for ease of presentation).  The square on
the left shows the median DIMM seeing, plotted at the DIMM elevation of 7~m.  Only DIMM and MASS data that were taken simultaneously with the SODAR data are used for this plot.
The curves are the sums of MASS seeing, XFAS SODAR seeing from 200 to 500~m and
the SFAS SODAR seeing from the height given by the abscissa to 200~m. 
There are two curves given for Armazones as two SFASs were working there
simultaneously for some time.  The good agreement between the DIMM seeing
and the sum of SFAS, XFAS and MASS seeing is evidence that our SODAR results are calibrated at the $<$10\% level for all sites except 
San Pedro M\'artir.  The discrepancy for SPM is caused by the noise
created by the trees at the site (wind noise as well as echos).  We nevertheless believe that the shapes of the curves are a good description of the average ground layer profiles for all sites, including San Pedro M\'artir. See TST-6 and references therein for details.

It must be noted that the SODAR data do not cover representative amounts of
time for all seasons at all sites.  Thus, the curves shown here should not be taken as representative for the long-term conditions at the sites. Instead, the main use of these curves is to obtain an estimate for the seeing that  TMT would encounter at any of the sites.  This seeing is, obviously, smaller than the
seeing measured by the DIMM at 7~m above the ground.  We can see that, at the top of the TMT enclosure, 50--60~m above the ground, the seeing differences between the sites are greatly reduced.  An exact quantitative analysis of this effect is not possible in a representative way as we do not have sufficient SODAR data.  Semi-quantitative estimates of its magnitude are used in the comparison of the TMT candidate sites in TST-12.  Finally, we point out that the presence of trees at San Pedro M\'artir raises the "effective level" of the ground for atmospheric turbulence.  In this sense, the SPM DIMM telescope is located less than 7~m above this "effective ground'', which accounts at least in part for the higher ground layer turbulence measured there.

Figure~\ref{f:cn2color} shows the median turbulence profiles as calculated from
the DIMM and MASS data.  The lowest data point of each profile (shown
here at 250~m) is the ground layer strength calculated from the
difference between the DIMM and MASS seeing.  The other points are the profiles
as measured by the MASS.  The plot shows again the stronger ground layer at
Mauna Kea 13N and San Pedro M\'artir, the stronger high-elevation turbulence of the Chilean sites and the
difference in the 16-km layer between Mauna Kea 13N and the other four sites,
resulting in the larger isoplanatic angle at Mauna Kea 13N.

\begin{table}[!t]
\begin{center}
{\scriptsize
\begin{tabular}{|l|c|c|ccccc|}
\hline
Parameter & Instrument & TST & Tolar & Armazones & Tolonchar & SP M\'artir & Mauna Kea 13N \\
\hline
\hline
Elevation [m] & & & 2290 & 3064 & 4480 & 2830 & 4050\\
\hline
Total seeing [as] & DIMM & 5 & 0.63 & 0.64 & 0.64 & 0.79 & 0.75\\
10\% DIMM seeing [as] & DIMM & 5 & 0.42 & 0.41 & 0.44 & 0.50 & 0.46\\
Free atmosphere seeing [as] & MASS & 5,6 & 0.44 & 0.43 & 0.48 & 0.37 & 0.33\\
10\% MASS seeing [as] & MASS & 5,6 & 0.24 & 0.23 & 0.25 & 0.17 & 0.15 \\
GL seeing  7~--~500~m [as] & D-M & 5,6 & 0.34 & 0.35 & 0.32 & 0.58 & 0.54 \\
Isoplanatic angle, $\theta_0$ [as] & MASS & 5,6 & 1.93 & 2.04 & 1.83 & 2.03 & 2.69\\
Coherence time, $\tau_0$ [ms] & M+D & 7 &  5.2 & 4.6 & 5.6 & 4.2 & 5.1 \\
\hline
Night temperature 2~m [$^\circ$C] & AWS & 8 & 14.0 & 7.5 & -0.7 & 5.4 & 2.3\\
Night wind speed 2~m [m/s] & AWS & 8 & 3.2 & 6.3 & 2.7 & (2.2) & 3.7\\
Night wind speed 7~m [m/s] & Sonic & 8 & 4.8 & 7.2 & 4.3 & (3.3) & 5.7\\
Night humidity 2~m [\%] & AWS & 8 & 19 & 21 & 36 & 38 & 30\\
T variation (10~--~90\%) [$^\circ$C] & AWS & 8 & 5.6 & 7.5 & 9.5 & 16.2 & 6.8\\
\hline
Clouds: clear fraction & Satellite & 9 & 87\% & 89\% & 82\% & 83\% & 76\% \\
PWV [mm] & Combination & 10 & 4.0 & 2.9 & 1.7 & 2.6 & 1.9\\
Fraction of PWV $<2$ mm & Combination & 10 & 18\% & 29\% & 62\% & 35\% & 54\%\\
\hline
\end{tabular}

\caption{\label{t:results-summary}Summary of results from the TMT candidate
sites.  These values are extracted from the more detailed results in the other papers in the TST series, as indicated in the third column.  All values are
medians, unless noted otherwise.  The expected uncertainties of the results are given in
TST-3 -- TST-11.  Data that are known to have problems are
shown in parentheses.}

}
\end{center}
\end{table}

Numerical values for some of the main site
characteristics shown in Figs.\ \ref{f:cumulcolor} -- \ref{f:cn2color} are given in
Table~\ref{t:results-summary}, along with the site elevations (given in Row 2 for reference) and the numbers of the papers of the TST series (Column 3) in which more details can be found:

\medskip

\noindent{\bf Total seeing, free atmosphere seeing:} Integrated seeing from the DIMM
and MASS, respectively.  Medians and best 10 percentile values are given.

\noindent{\bf Ground layer (GL) seeing 7~--~500~m:} Ground layer seeing calculated from
the difference between the DIMM and MASS.  The GL seeing as
measured by the SFAS SODAR is not given here because the SODAR data do not cover
representative periods for all sites (see Fig.~\ref{f:seeing_height}).

\noindent{\bf Isoplanatic angle:} Isoplanatic angle, $\theta_0$, obtained from MASS profiles.

\noindent{\bf Nighttime temperature, wind speed and humidity:} Meteorological values
obtained at 2~m above the ground with the weather stations and at 7~m with the
sonic anemometer.

\noindent{\bf Temperature variations (10~--~90\%):} Difference between the
ninetieth and tenth percentile of the temperature distributions measured at 2~m above the ground, thus giving
the temperature range not exceeded 80\% of the time.

In addition, the table also lists a number of other key parameters of the TMT candidate sites:

\noindent{\bf Turbulence coherence time:} The turbulence coherence time, $\tau_0$, is obtained from combining the MASS (free atmosphere) and ground layer $\tau_0$ measurements.  The MASS $\tau_0$ is calculated from the temporal variations of stellar scintillations.  The GL $\tau_0$ is calculated from the low elevation wind in combination with the GL turbulence strength.  These values are extrapolations of the coherence times that TMT would encounter 50~--~60~m above the ground.  While the contributions of GL and free atmosphere to $\tau_0$ can be very different from site to site, it can be seen that the total coherence times are similar for all sites.  See TST-7 for details.

\noindent{\bf Cloud cover, clear fraction:} The cloud-free fractions of time are obtained from satellite data, as it is very important to have the longest possible temporal baseline available for this analysis.  They have been verified to be consistent with ground based ASCA photometry and MASS transparency data for simultaneous periods at the 3\% level for all sites except Tolonchar, for which the agreement is at the 6\% level.  Note that a very specific definition of "cloud free" fraction is used here due to the method used.  It should not be confused with the the fraction of photometric nights or the time an observatory can be used at the sites.  The definition is, however, consistent between the sites.  The results show the expected behavior, the extremely cloud-free conditions of the western Atacama desert (Tolar and Armazones), with somewhat more clouds encountered at Tolonchar and San Pedro M\'artir, followed by Mauna Kea 13N.   Details are discussed in TST-9, which also provides a discussion of the expected time lost in addition to clouds due to other inclement weather conditions such as high wind or humidity.

\noindent{\bf PWV and fraction of PWV $<2$~mm:}  The last two rows of Table \ref{t:results-summary} show the median precipitable water vapor (PWV) value and the fraction of time the PWV value is below 2~mm, obtained from a combination of ground-based (IRMA and other radiometers) and satellite data (see TST-10).  It shows the expected behavior of PWV being an exponentially decreasing function of altitude.  The only exception to this rule of thumb is San Pedro M\'artir, which is a little drier than would be expected based on its altitude alone.

As a final remark, we reiterate that TMT site selection will not be based on the exact values of any single parameter at any of the sites.  It is based on a careful balancing act of atmospheric, technical and programmatic parameters in which small changes in one parameter are unlikely to affect the outcome.  For example, while we believe that our DIMMs measure seeing with a reproducibility significantly better than 0.05 arcsec (and probably with an accuracy of the same order), differences between sites of that order or even more will, by themselves, not eliminate a site from the list of sites that are qualified for TMT.  It has always been understood that annual and multi-year variations can easily be on this order and, furthermore, that the seeing that TMT will encounter during its lifetime is different from the measured DIMM seeing anyway, due to the telescope's elevation above the ground, modifications made to the terrain in the construction of such a large structure (which will be different for each site) and the outer scale of turbulence (which might also be different from site to site).  The same kind of considerations apply to all other parameters presented here.  In the end, the site testing data serve the purpose of identifying which sites are {\it qualified} for hosting TMT as far as their atmospheric properties are concerned.  The process by which the sites' qualifications are assessed is presented in TST-12.

\section{Summary}

Over the last eight years, the Thirty Meter Telescope and its precursor projects have invested considerable resources into identifying the best possible candidate sites for TMT.  The work began by considering as complete a list as feasible of all potentially interesting sites on Earth.  This list was first narrowed down to three regions in the western hemisphere, northern Chile, southwestern North America, and Hawaii based on existing information from previous site testing work and general knowledge about the atmosphere.  Through analyses of cloud cover and precipitable water vapor using satellite data, five candidate sites were selected for in-situ testing, namely Cerros Tolar, Armazones and Tolonchar in northern Chile; San Pedro M\'artir in northern Baja California, Mexico and Mauna Kea 13N, Hawaii.

TMT then began an extensive on-site testing effort for which we equipped each of the five candidate sites with a large array of instrumentation for analyzing atmospheric properties, with a special emphasis on turbulence and turbulence profiles.  All equipment was calibrated and characterized carefully before deployment to the candidate sites, where it operated for 2.5 to 5 years.  The results show that all candidate sites are excellent and are clearly among the best ground-based telescope sites on Earth.  They also show that not a single site is perfect or is the best (or worst) in all parameters.  This is, of course, a welcome outcome, as it means that the selection of candidate sites was successful and that the final site selection can safely also take into account other considerations, without compromising the expected scientific output of TMT.

Based on the site testing data and on other technical and programmatic considerations, the TMT Project selected two sites for further consideration, one in the northern hemisphere, one in the southern hemisphere.  They are Cerro Armazones and Mauna Kea 13N.

\section*{Acknowledgments}

We acknowledge the contributions of three of our friends and collaborators who did not live long enough to see the results of this work: D. Andr\'e Erasmus, Juan Araya and Hugo E. Schwarz.  The TMT site selection work could not have happened without the direct contributions of almost 100 people -- acknowledged by name in the `Contributors' section of the TMT site testing website -- at various institutions.  A special thanks to everybody who has
supported the TMT site testing program, in
particular the people at the Cerro Tololo Inter-American Observatory (CTIO),
the National Optical Astronomy Observatory (NOAO) and the AURA New Initiatives
Office (NIO), the Universidad Cat\'olica del Norte in Antofagasta, the
Observatorio Astron\'omico Nacional de San Pedro M\'artir, Gemini
Observatory,   Palomar Observatory, the Submillimeter Array (SMA), the
University of Hawaii Institute for Astronomy (IfA), the University of
Lethbridge, the University of Moscow, the South African Astronomical
Observatory (SAAO), and the entire TMT project including its partner
institutions and external reviewers.

The TMT Project gratefully acknowledges the support of the TMT partner institutions. They are the Association of Canadian Universities for Research in Astronomy (ACURA), the California Institute of Technology and the University of California. This work was supported as well by the Gordon and Betty Moore Foundation, the Canada Foundation for Innovation, the Ontario Ministry of Research and Innovation, the National Research Council of Canada, the Natural Sciences and Engineering Research Council of Canada, the British Columbia Knowledge Development Fund, the Association of Universities for Research in Astronomy (AURA) and the U.S. National Science Foundation.





\end{document}